\documentclass[twocolumn,nobibnotes,superscriptaddress,showpacs, aps, prd]{revtex4-1}
\usepackage{graphicx}
\usepackage{amsmath,amssymb}
\usepackage{booktabs}
\usepackage{subfigure}
\usepackage{tabularx}
\usepackage{url}
\usepackage{ulem}

\makeatletter
\renewcommand*{\@thesubfigure}{\thesubfigure\space}
\renewcommand*{\@thesubtable}{\thesubtable\space}
\makeatother

\newcolumntype{C}{>{\centering\arraybackslash}X}
\newcolumntype{R}{>{\raggedleft\arraybackslash}X}
\newcolumntype{L}{>{\raggedright\arraybackslash}X}

\newcommand{\rme}{\mathrm{e}}
\newcommand{\rmi}{\mathrm{i}}
\newcommand{\rmd}{\mathrm{d}}
\newcommand{\omegaI}{\omega_\mathrm{I}}
\newcommand{\omegaR}{\omega_\mathrm{R}}

\DeclareMathOperator*{\argmin}{argmin}
\newcommand{\Err}{\mathrm{Err}}

\begin{document}

\title{Estimation of starting times of quasinormal modes in ringdown gravitational waves with the Hilbert-Huang transform}

\author{Kazuki Sakai}%
\email{k\_sakai@stn.nagaokaut.ac.jp}
\affiliation{Department of Information Science and Control Engineering, Nagaoka University of Technology, Niigata 940-2188, Japan}
\author{Ken-ichi Oohara}
\affiliation{Graduate School of Science and Technology, Niigata University, Niigata 950-2181, Japan}
\author{Hiroyuki Nakano}
\affiliation{Faculty of Law, Ryukoku University, Kyoto 612-8577, Japan}
\affiliation{Department of Physics, Kyoto University, Kyoto 606-8502, Japan}
\author{Masato Kaneyama}
\affiliation{Department of Physics, Graduate School of Science, Osaka City University, Osaka 558-8585, Japan}
\author{Hirotaka Takahashi}
\affiliation{Department of Information and Management Systems Engineering, Nagaoka University of Technology, Niigata 940-2188, Japan}
\affiliation{Earthquake Research Institute, The University of Tokyo, Tokyo 113-0032, Japan}

\date{\today}

\begin{abstract}
It is known that a quasinormal mode (QNM) of a remnant black hole dominates a ringdown gravitational wave (GW)
    in a binary black hole (BBH) merger.
To study properties of the QNMs,
    it is important to determine the time when the QNMs
    appear in a GW signal 
    as well as to calculate its frequency and amplitude.
In this paper,
    we propose a new method of estimating the
    starting time of the QNM
    and calculating the QNM frequency and amplitude of BBH GWs.
    We apply it to simulated merger waveforms by numerical relativity and the observed data of GW150914.
    The results show that
    the obtained QNM frequencies and time evolutions of amplitudes
        are consistent with the theoretical values within 1\% accuracy
        for pure waveforms free from detector noise.
In addition,
    it is revealed that there is a correlation between the
      starting time of the QNM and the spin of the remnant black hole.
In the analysis of GW150914, we show that the
    parameters of the remnant black hole estimated through our method
    are consistent with those given by LIGO
    and a reasonable starting time of the QNM is determined.
\end{abstract}

\maketitle

\section{Introduction}

Advanced LIGO detected two events of gravitational waves (GWs) in 2015,
    both of which are generated by binary black hole (BBH) mergers.
The waveforms of GWs from BBHs consist of three phases,
    namely, the inspiral, merger and ringdown phases.
The inspiral phase
    corresponds to an orbital motion of the two black holes (BHs)
    and can be described by the post-Newtonian approximation of general relativity (GR).
The merger phase is described as a full nonlinear evolution of Einstein equations
    and derived by numerical relativity simulations.
The ringdown phase corresponds to quasinormal modes (QNMs) of the remnant BH\@.
Since the QNMs are intrinsic in the BH metric and can be seen only with GWs,
    analyses of the QNMs with GWs enable us to test GR at a strong and dynamical gravitational field.

A QNM is expressed as a damped sinusoid
\begin{align}
    h_\mathrm{QNM}(t)
        &= \Re \left[ A_0 \rme^{-\rmi \{ \omega (t - t_0) + \varphi_0 \}} \right]
      \notag \\
        &= A_0 \rme^{\omegaI (t - t_0)} \cos \bigl( \omegaR (t - t_0) + \varphi_0 \bigr),
      \label{eq:hQNM}
\end{align}
where $\omega = \omegaR + \rmi \omegaI$ is the QNM frequency,
    and $A_0$, $t_0$, $\varphi_0$ are the initial amplitude, time, and phase, respectively.
Nakano \textit{et al}.~\cite{Nakano_PRD_2015} have proposed a method to test GR using observed QNM GWs.
While the waveform in the ringdown phase consists of the sum of an infinite number of QNMs,
    they focused on the dominant, slowest-damped mode [$(l,m,n) = (2,2,0)$]
    and revealed that
    the QNM frequency can be estimated with an accuracy at the $5 \sigma$ level
        for a typical case of the mass and spin of the remnant black hole
        if the signal-to-noise ratio (SNR) of observed GWs is sufficiently high~\cite{Kinugawa_PTEP_2016}.
It means that we can conclude GR is inconsistent with the event
    if the QNM frequency estimated from it takes a value in the region forbidden by GR\@.
This method can determine
    how high SNR for the QNM signal is required to test GR for each mass of BHs.
We need to estimate the starting time of the QNM in the ringdown phase
    to precisely estimate the QNM frequency
    and to calculate the SNR for the QNM from observed GWs from BBHs.
According to Nakano \textit{et al}.~\cite{Nakano_PRD_2015},
    the roughly estimated rate of detections of BBHs with SNR $> 50$
        by the second generation GW detectors,
        such as Advanced LIGO~\cite{Abbott_LRR_2016,Aasi_LIGO},
        Advanced Virgo~\cite{Abbott_LRR_2016,Acernese_Virgo},
        and KAGRA~\cite{Kuroda_KAGRA1,Aso_KAGRA2},
        is 2 yr$^{-1}$.
To make the best of such chances to test GR,
    establishing a method of estimating the starting time is important.
However,
    it is difficult to phenomenologically define the starting time,
    because the transition from the merger to ringdown phase still has not been well understood.
In a theoretical approach,
    effects of the determination of the starting time
        for residuals in the fitting of QNM signals and for extractions of parameters
    have been investigated by using only QNM signals~\cite{Dorband_PRD_2006}
        and by using full numerical-relativity waveforms of BBH~\cite{Berti4_PRD_2007,Berti_PRD_2007}.
In this paper,
    we propose a method to estimate the starting time
        in a signal-processing approach
        by using the Hilbert-Huang transform (HHT).

Since the Hilbert-Huang transform is not limited by the time-frequency uncertainty relation,
    unlike the short-time Fourier transform (STFT) and the wavelet transform (WT),
    it provides high time-frequency resolution,
      and therefore enables us to investigate phenomena that have even rapid changes in frequency
        as well as no or slow changes, while the STFT and the WT can detect only the latter.
Application of the HHT to GW data analysis has been examined with various ways~\cite{Jordan_PRD_2007,Stroeer_PRD_2009,Takahashi_AADA_2013,Kaneyama_PRD_2016,Sakai_ICICEL_2017}.

The HHT consists of two steps:
    a mode decomposition step and a spectral analysis step.
At the mode decomposition step,
    input data will be decomposed into some intrinsic mode functions (IMFs).
Then, at the spectral analysis step,
    the time evolution of the amplitude and the phase of each IMF are calculated by means of Hilbert transform.
This spectral analysis is called Hilbert spectral analysis (HSA).
Hence, we can express input data $s(t)$ as
\begin{align}
    s(t)
        &= \sum_{i=1}^{N_\mathrm{IMF}} c_i(t) + r(t)
      \label{eq:HHT1} \\
        &= \sum_{i=1}^{N_\mathrm{IMF}} a_i(t) \cos ( \phi_i(t) ) + r(t),
      \label{eq:HHT2}
\end{align}
where $N_\mathrm{IMF}$ is the number of IMFs,
and $c_i(t)$, $a_i(t)$, $\phi_i(t)$ are
    the $i$th IMF, the instantaneous amplitude (IA), and the instantaneous phase (IP) respectively,
    while $r(t)$ is the residual or trend, which is the nonoscillatory part mode of the data.
Instantaneous frequency (IF) $f_i(t)$ can be defined as
\begin{align}
    f_i(t) = \frac{1}{2 \pi} \frac{\rmd \phi_i(t)}{\rmd t}.
\end{align}

The original HHT (proposed in \cite{Huang_PRSL_1998}) uses the empirical mode decomposition (EMD) as a decomposition method.
It is known, however,
    that the EMD has many drawbacks, such as
    mode-mixing~\cite{Datig_OE_2004,Rilling_IEEE_2008}, mode-splitting~\cite{Yeh_AADA_2010}, and lack of mathematical foundation~\cite{Colominas_DSP_2015},
    and therefore many improved methods of the EMD have been being proposed~\cite{Meignen_IEEE_2007,Hong_IEEE_2009,Huang_JCAM_2013,Colominas_DSP_2015,Yeh_AADA_2010,Colominas_BSPC_2014}.
We will adopt the ensemble EMD (EEMD),
    which is proposed by Z.~Wu and N.~E.~Huang~\cite{Wu_AADA_2009} to overcome the mode-mixing problem.
The detail of the EEMD is also written in~\cite{Kaneyama_PRD_2016}.

In this paper,
  we propose a method of 
  estimating the starting time of the QNM in the ringdown phase of BBH GW,
  and report the results of evaluation for the efficiency of the method.
Following Nakano \textit{et al.}~\cite{Nakano_PRD_2015},
    we will focus on the dominant, slowest-damped mode [$(l,m,n)=(2,2,0)$].
It is realized by applying a bandpass filter to each waveform before the HHT
    and choosing appropriate parameters of the EMD,
    as described in Sec.~\ref{sec:apply_filter_HHT} below.
The basic concepts and algorithm of the method are described in Sec.~\ref{sec:strategy},
  and the setups and the results of simulation for the evaluation are shown in Sec.~\ref{sec:simulation}.
In Sec.~\ref{sec:GW150914}, 
  the result of application of the method to the observed data of GW150914 is described.
Section \ref{sec:summary} is devoted to a summary.

Throughout this paper, discrete sequences are represented with brackets, such as $x[n]$,
and continuous functions are represented with parentheses, such as $s(t)$.
The Fourier transform of a continuous function $s(t)$
    is represented as $\tilde{s}(f)$.

We perform calculation of the HHT using KAGRA Algorithmic Library (KAGALI)~\cite{KAGALI_MG_2017}.

\section{Strategy of our method}
\label{sec:strategy}

We assume it is already known that
    the data contains a GW signal from a BBH merger.

If the QNM is perfectly extracted in the $j$th IMF,
    it follows from Eqs.~\eqref{eq:hQNM} and \eqref{eq:HHT2}
        that the IA and IP of the IMF are give by
\begin{align}
    a_j(t)
        &= A_0 \rme^{\omegaI(t - t_0)},
      \label{eq:IA_fits} \\
    \phi_j(t)
        &= \omegaR(t - t_0) + \varphi_0.
      \label{eq:IP_fits}
\end{align}
In reality,
    the IMF contains also other modes before the QNM starts,
    and noise components become dominant after the QNM is sufficiently damped.
Equations~\eqref{eq:IA_fits} and \eqref{eq:IP_fits} do not hold in the merger phase and the noise-dominant segment.
If the IA fits Eq.~\eqref{eq:IA_fits} most properly during a significant period from a certain time,
    it can be defined as the starting time of the QNM\@,
    and we call the segment a \textit{QNM-dominant segment}.

\begin{figure}[tbp]
    \centering
    \includegraphics[clip]{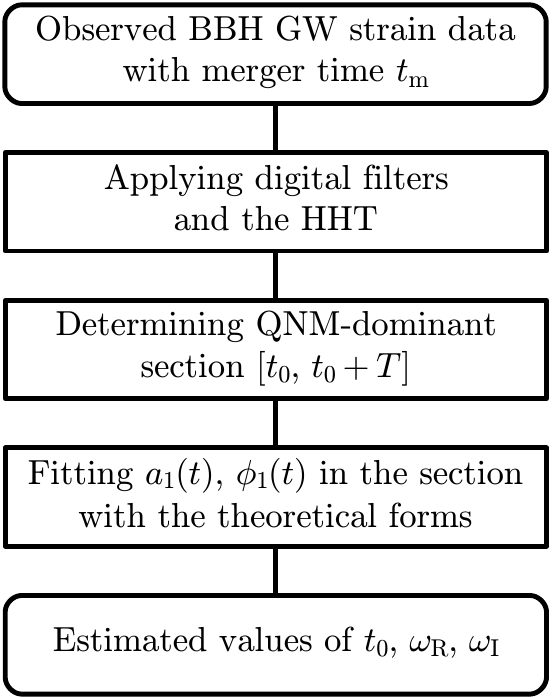}
    \caption{Overall flowchart of estimation of QNM parameters.}
    \label{fig:flow_main}
\end{figure}

The general flowchart of our proposed method is shown in Fig.~\ref{fig:flow_main}.
The method consists of three steps:
    (A) applying digital filters and the HHT,
    (B) determining a QNM-dominant segment,
    (C) and fitting IA and IP of the IMF of interest with the theoretical forms.

\subsection{Applying digital filters and the HHT}
\label{sec:apply_filter_HHT}

Raw strain data contain detector noise generated from various origins,
    such as seismic noise, quantum shot noise, suspension resonance, and so on.
Some of them are relatively strong and narrow band noises,
    called line noises.
As shown in Ref.~\cite{LIGO_GW150914_detection_PRL},
    many line noises exist at the Advanced LIGO's first observing run (O1).

In this step,
    to extract a sensitive band and to attenuate line noises from raw data,
    some digital filters are applied to the data.
For extraction of the sensitive band,
    one bandpass filter will be applied,
  and for attenuation of each line noise,
    a corresponding notch filter will be applied.
After the data processed by the filters,
    the HHT is applied in order to extract the QNM signal.

We must determine some parameters of the filters and the HHT\@.
The lower cutoff frequency of the bandpass filter should be equal to the lower limit of the properly calibrated band.
For other parameters, however,
    since there is not a universal principle applicable in various situations,
    they should be determined based on each targeted data.
In our method,
    we attempt to extract the QNM of $(l,m,n)=(2,2,0)$ into the IMF1,
        which contains the highest-frequency components,
        in order to prevent the other IMFs from absorbing some components of the target mode.
Wu and Huang~\cite{Wu_PRSL_2004} revealed that IMF1 sometimes contains a small but unignorable fraction of components
    in frequency bands of other IMFs.
Therefore,
    if we attempt to extract a target mode into the second or higher IMF,
    some fraction of the component may be absorbed into the IMF1.
To extract the QNM into the IMF1,
    we choose the parameters that maximize the signal energy of the IMF1
        in the segment containing the merger phase and the beginning of the ringdown phase.
The segment will be represented as $[t_\mathrm{m},\,t_\mathrm{m}+T_\mathrm{m}]$,
    if $t_\mathrm{m}$ denotes the time when the amplitude of the GWs is at its maximum
    and $T_\mathrm{m}$ denotes a certain duration that covers the merger phase
    and the beginning of the ringdown phase.
In this paper,
    we determined to fix $T_\mathrm{m}$ at 10 ms for all input data after several checks.
We have confirmed that this value of $T_\mathrm{m}$ does not cause any trouble for our tests,
    but if we consider larger total mass than $\sim 60 M_\odot$, $T_\mathrm{m}$ should take a greater value.
We choose the upper cutoff frequency of the bandpass filter
    to make the dominant $(l,m)=(2,2)$ mode be the highest frequency components of the filtered data.
The nondominant higher modes, such as $(l,m) = (3,3)$ mode, will be eliminated in this step.

\begin{figure}[htbp]
    \centering
    \includegraphics[clip]{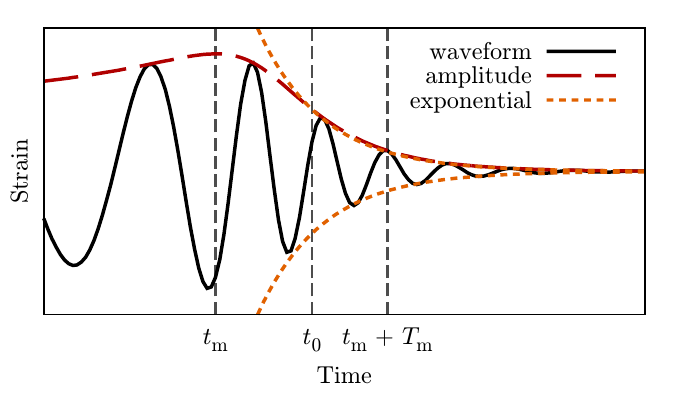}
    \caption{
        Illustration of the time parameters,
            where $t_\mathrm{m}$ is the time when the amplitude is at its maximum,
            $t_0$ is the starting time of the QNM,
            and $T_\mathrm{m}$ is a certain duration
                that covers the merger phase and the beginning of the ringdown phase.
        They are shown along with a strain of BBH GW (black solid line),
            its amplitude (red dashed line),
            and the best-fitted exponential curve of the QNM (orange dotted line).
        In the preprocessing step,
            the parameters will be chosen 
                to maximize the signal power of the IMF1 for the segment $[t_\mathrm{m}, t_\mathrm{m}+T_\mathrm{m}]$.
        In this paper,
            $T_\mathrm{m}$ is fixed at 10 ms.
    }
    \label{fig:times}
\end{figure}

\subsection{Determining QNM-dominant segment}

\begin{figure}[tbp]
    \centering
    \includegraphics[clip]{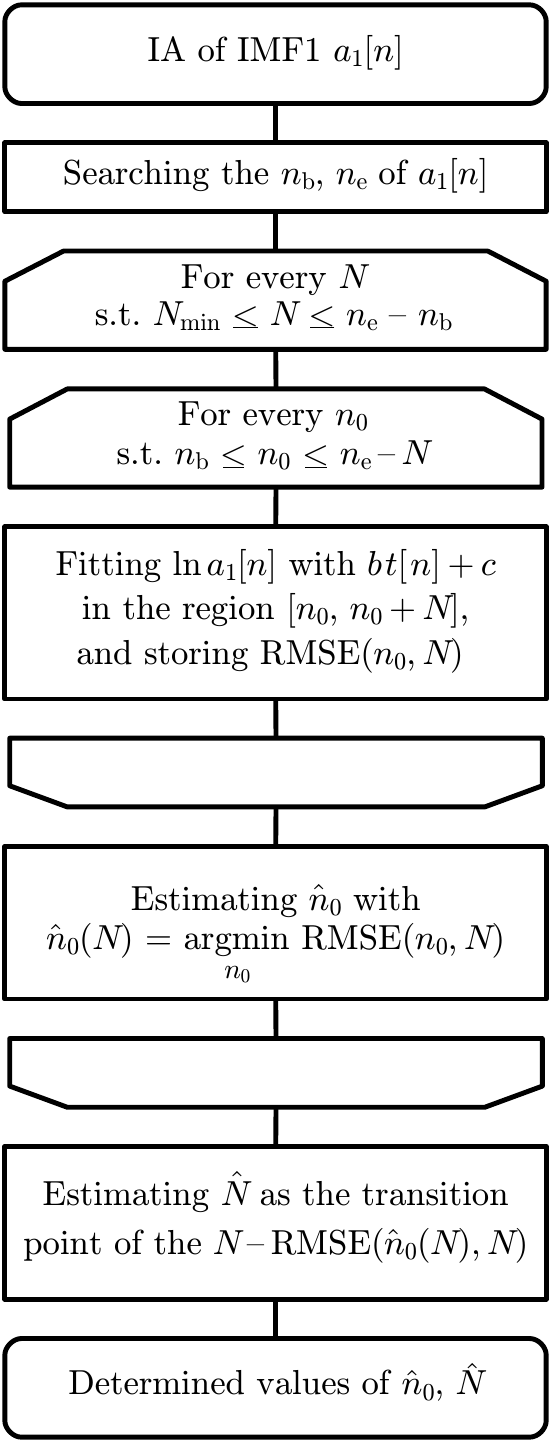}
    \caption{Flowchart of determining the QNM-dominant segment.
        The indices $n_\mathrm{b}$ and $n_\mathrm{e}$ respectively denote
            the starting time and the ending time of the longest segment where the IA, $a_1[n]$, decreases monotonically.
        The index $\hat{n}_0$ of the starting time of the QNM and
            the duration $\hat{N}$ of the QNM-dominant segment
            are determined from the exponentially decaying curve best-fitted to the IA.
    }
    \label{fig:flow_fitting1}
\end{figure}

After the QNM signal is extracted into the IMF1,
    we will estimate the starting time of the QNM $\hat{t}_0$ and the duration of QNM-dominant segment $\hat{T}$.
Figure~\ref{fig:flow_fitting1} shows the flowchart of determining the segment.

\begin{figure}[htbp]
    \centering
    \includegraphics[width=7cm]{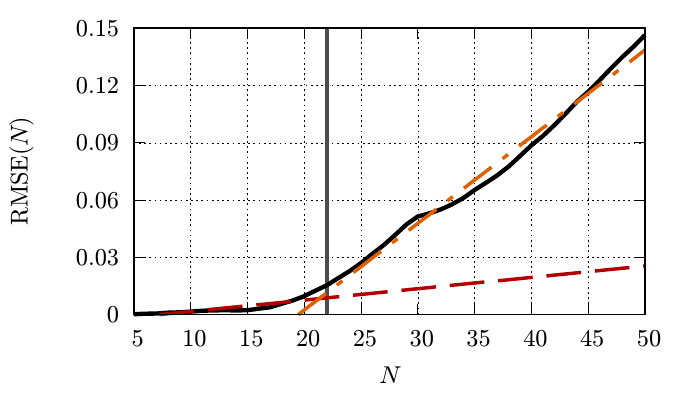}
    \caption{
        How to determine $\hat{N}$.
        The black solid line shows $\mathrm{RMSE}(\hat{n}_0(N),N)$ we obtain.
        The red dashed line and the orange dot-dashed line illustrate the linear regression of it
            for small and large values of $N$, respectively.
        In this case,
            $\hat{N}$ is determined to be 22, which is shown by the gray vertical line.
    }
    \label{fig:oresen}
\end{figure}

The IA in the QNM-dominant segment decreases monotonically,
    but monotonically decreasing segments are not always the QNM-dominant segment.
This means that we need to determine the QNM-dominant segment in these segments.
To begin with,
    we search for the longest segment where the IA decreases monotonically.
We assume that it extends from $t[n_\mathrm{b}]$ to $t[n_\mathrm{e}]$,
    or in the interval $[n_\mathrm{b}, n_\mathrm{e}]$.
And then we make a linear regression of the logarithm of IA, $\ln a_1[n]$,
    on a linear function $b t[n] + c$ for every possible subsegment $[n_0,\,n_0 + N]$,
    where $N_\mathrm{min} \leq N \leq n_\mathrm{e} - n_\mathrm{b}$
        and $n_\mathrm{b} \leq n_0 \leq n_\mathrm{e} - N$,
        with $N_\mathrm{min}$ being a predetermined minimum size of the fitting interval.
There is no definite principle of determining $N_\mathrm{min}$,
    but we set it to 5 here.
Defining the root mean squared error (RMSE) of the fitting as
\begin{align}
    \mathrm{RMSE}(n_0, N)
        = \sqrt{ \frac{1}{N} \sum_{n = n_0}^{n_0 + N - 1} \left( \ln a_1[n] - bt[n] - c \right)^2 },
\end{align}
$\hat{n}_0(N)$,
    which denotes the optimal value of $n_0$ for each $N$,
    is determined by
\begin{align}
    \hat{n}_0(N) = \argmin_{n_0} \mathrm{RMSE}(n_0, N).
\end{align}
Finally,
    $\hat{N}$, which is the optimal value of $N$, is determined as the transition point of a slope of $N$ -- $\mathrm{RMSE}(\hat{n}_0(N),N)$ plot,
    based on following two hypotheses:
\begin{itemize}
    \item   If the segment $[\hat{n}_0(N),\, \hat{n}_0(N) + N]$ is a part of the QNM-dominant segment,
            RMSE gradually increases with $N$,
            because only detector noise contributes to fitting error.
    \item   If the segment $[\hat{n}_0(N),\, \hat{n}_0(N) + N]$ contains other modes,
            a merger phase or noise-dominant segment,
            RMSE rapidly increases with $N$,
            because fitting basis is not proper for the modes.
\end{itemize}
Specifically,
    $\hat{N}$ is determined as
\begin{gather}
    \hat{N}
        = \argmin_{N} \left[ \Err(N_{\min}, N) + \Err(N + 1, N_{\max}) \right],
    \\
    \Err(N_1, N_2)
        = \min_{a,b} \sqrt{\frac{\sum_{N = N_1}^{N_2} \left( e(N) - (aN + b) \right)^2}{N_2 - N_1}},
    \\
    e(N)
        = \mathrm{RMSE}(\hat{n}_0(N), N).
\end{gather}
Figure~\ref{fig:oresen} illustrates how to determine $\hat{N}$.
It is plotted for the same data with Fig.~\ref{fig:example_IMF}\subref{fig:0262_IMF_good},
    which will be described in Sec.~\ref{sec:simulation} later.
In this case,
    $\hat{N}$ is determined to be $22$.

\subsection{Fitting IA and IP in the segment}

Since taking the logarithm of Eq.~\eqref{eq:IA_fits} gives
\begin{equation}
    \ln a_1(t) = \omegaI t - \omegaI t_0 + \ln A_0,
\end{equation}
we can estimate $\omegaI$ and $A_0$ 
    if $\ln a_1[n]$ is successfully fitted onto a linear regression function $b t[n] + c$ in a segment $[n_0,\, n_0 + N]$,
    such that
\begin{align}
    \hat{\omega}_\mathrm{I}
        &= b
      ,\quad
    \hat{A}_0
         = \exp(c + b t[n_0]).
\end{align}
Also, $\omegaR$ and $\varphi_0$ can be estimated by fitting $\phi_1[n]$ with $d t[n] + e$ as
\begin{align}
    \hat{\omega}_\mathrm{R}
        &= d
      ,\quad
    \hat{\varphi}_0
         = e + d t[\hat{n}_0].
\end{align}

Since GWs are redshifted,
    the QNM frequencies estimated here should be shifted by the factor $1/(1 + z)$,
    where $z$ is the cosmological redshift parameter of the source.
For tests of GR,
    frequencies should be evaluated at the source frame
        and compared with those estimated with the theory.
The values of frequencies given for the rest of the paper are
    what are evaluated at the source frame,
\begin{align}
    \hat{\omega}_\mathrm{R}^\mathrm{src.}
        &= (1 + z) \hat{\omega}_\mathrm{R}
      ,\quad
    \hat{\omega}_\mathrm{I}^\mathrm{src.}
         = (1 + z) \hat{\omega}_\mathrm{I}.
\end{align}
Hereinafter,
    they will be denoted simply by $\omegaR$ and $\omegaI$ without the superscript ``src''.

\section{Simulation}
\label{sec:simulation}

We conduct some simulations in this section to evaluate the efficiency of our method.
We make use of waveforms obtained through numerical relativity simulation of BBH mergers
    by the SXS (Simulating eXtreme Spacetimes) project~\cite{SXS_CQG_2016,SXS_HP}.
First,
    we make analysis of pure GWs free from detector noise
    and compare the obtained values of the QNM frequencies
        to the theoretical values calculated with the parameters given by the SXS\@.
And then,
    we make ``simulated observation data'' by injecting these waveforms to simulated detector noise
    and apply the procedure to them in order to simulate realistic analysis with observed strain data.

\begin{figure}[htbp]
    \centering
    \includegraphics{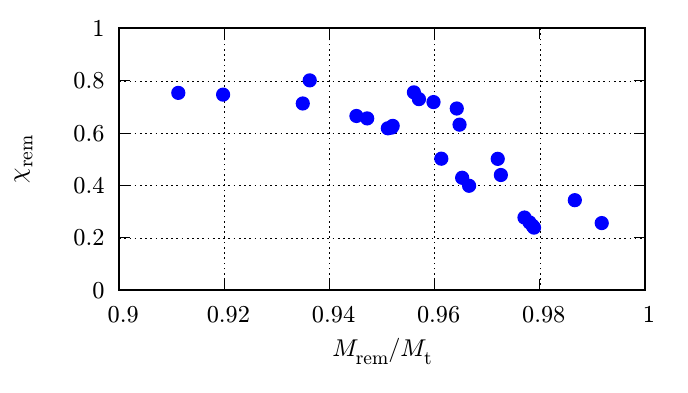}
    \caption{
        Parameter distributions of the remnant BHs in the SXS catalog for the waveforms that we use.
        The horizontal axis represents the ratio of the mass of the remnant BH $M_\mathrm{rem}$
            to the initial total mass of the BBH $M_\mathrm{t}$,
          and the vertical axis represents the dimensionless spin of the remnant BH $\chi_\mathrm{rem}$.
    }
    \label{fig:param_dist}
\end{figure}

The SXS project provides gravitational waveforms from BBH mergers
    in their web site~\cite{SXS_HP}.
After the detection of GW150914,
    96 waveforms newly presented
        to serve to validate and improve aligned-spin waveform models for GW science~\cite{SXS_CQG_2016}.
Each waveform is characterized by the mass ratio, $q = m_1/m_2$,
    and the initial dimensionless spins, $\vec{\chi}_1$ and $\vec{\chi}_2$,
    where $m_1$ and $\vec{\chi}_1$ are the mass and the initial spin of the primary BH of the BBH, respectively,
    while $m_2$ and $\vec{\chi}_2$ are those of the secondary BH\@.
The waveforms can be classified into several categories:
    Neither is spinning, only one is spinning, and both are spinning.
The direction of the spins may be aligned or antialigned with the orbital angular momentum.
The magnitude of the spins may be equal or different if both are spinning.
We select some waveforms from each category
    such that the largest mass ratio case, the largest spin magnitude case, and so on are included.
As a result, we selected a total of 24 waveforms out of the 96 waveforms.
Moreover,
    we add two waveforms, which best fit GW150914 and GW151226.
The distribution of remnant BH's masses and spins is shown in Fig.~\ref{fig:param_dist}.

According to Kinugawa \textit{et al}.~\cite{Kinugawa_MNRAS_2014},
    the mass distribution of population III BHs has a peak at $30 M_\odot$,
    and for this reason, we set the total mass of a BBH
    to $60 M_\odot$.
We assume that the BBH is located at $z = 0.09$,
    as GW150914 and GW151226 are,
    and adopt a flat $\Lambda$CDM cosmology with the cosmological parameters observed by the Planck project~\cite{Planck_AANDA_2016},
        where Hubble parameter $H_0 = 67.9$ $\mathrm{km\;s^{-1}\;Mpc^{-1}}$
        and matter density parameter $\varOmega_\mathrm{m} = 0.306$.

In order to make simulated observation data,
    we generate Gaussian noises based on Advanced LIGO's design sensitivity,
        the zero-detuned high-power sensitivity curve~\cite{aLIGO_ZDHP},
    and add simulated waveforms to them.

\begin{figure*}[tbp]
    \hfill
    \subfigure[SXS:BBH:0305]{%
        \includegraphics[width=.32\textwidth]{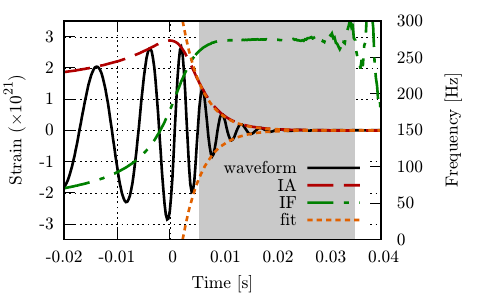}%
        \label{fig:0305_fit}%
    }
    \hfill
    \subfigure[SXS:BBH:0262]{%
        \includegraphics[width=.32\textwidth]{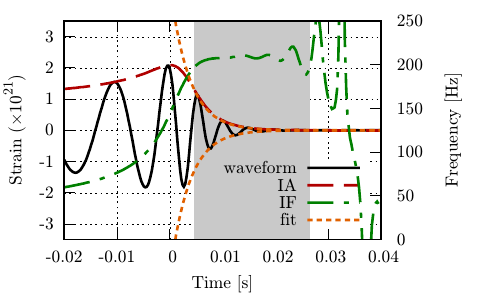}%
        \label{fig:0262_fit}%
    }
    \hfill
    \subfigure[SXS:BBH:0230]{%
        \includegraphics[width=.32\textwidth]{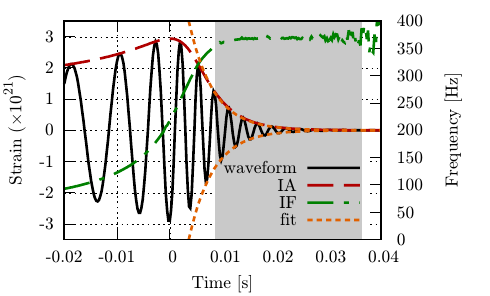}%
        \label{fig:0230_fit}%
    }
    \hfill\null
    \caption{
        Fitting of noise-free waveforms.
        As representative examples,
            results obtained for SXS:BBH:0305, 0262 and 0232 are shown.
        In each panel,
            the black solid line shows the waveform analyzed,
            the red dashed line and the green dot-dashed line are the IA and the IF
                calculated with the Hilbert spectral analysis, respectively.
        The QNM-dominant segment we determined is represented by the area filled with the color gray,
            although only the horizontal time span of the area is meaningful.
        The upper and lower orange dotted lines show the exponentially decaying curve
            fitted to the IA and its sign-inversion, respectively.
    }
    \label{fig:ideal_fits}
\end{figure*}

To begin with,
    we apply our method to noise-free gravitational waveforms.
Figure~\ref{fig:ideal_fits} illustrates how the method works.
The blue solid line shows the waveform analyzed,
    the red dashed line and the green dot-dashed line are the IA and the IF calculated with the Hilbert spectral analysis, respectively.
The QNM-dominant segment we determined is represented by the area filled with the color gray,
    although only the horizontal time span of the area is meaningful.
The upper and lower orange dotted lines show the exponentially decaying curve fitted to the IA and its sign-inversion, respectively.
It is apparent that the IF becomes invariant with time in the QNM-dominant segment as expected from Eq.~\eqref{eq:hQNM},
    while the IF oscillates in the latter part of this segment.
The oscillation is caused by the fact that
    the amplitude becomes very small, and therefore numerical errors grow significant.

\begin{figure}[htbp]
    \centering
    \includegraphics[clip]{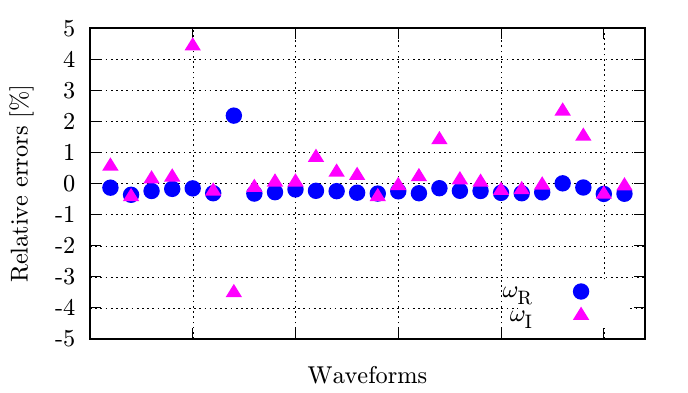}
    \caption{
        Relative errors between theoretical values of the QNM frequencies
            and values estimated through our analysis with noise-free simulated waveforms,
            i.e. $(\omegaR - \hat{\omega}_\mathrm{R})/\omegaR$ and $(\omegaI - \hat{\omega}_\mathrm{I})/\omegaI$.
        The blue circles and the magenta triangles represent the relative errors for $\omegaR$ and $\omegaI$, respectively.
    }
    \label{fig:omega_comparison}
\end{figure}

Figures \ref{fig:omega_comparison}
    represents the relative errors of the QNM frequencies
        between what we obtain and the theoretical values,
        which are calculated for the spin and the mass of the remnant BH
        by using the table of QNM frequencies against parameters of the BH
            released by Berti~\cite{Berti_PRD_2006,Berti_HP}.
The values of the spin and the mass are presented in the metadata of each simulation.
It is shown that the obtained frequencies agree with theoretical values sufficiently with some exceptions,
    while discrepancy in $\omegaI$ is somewhat larger than in $\omegaR$.
Some of the SXS waveforms in the ringdown phase look deformed,
    especially in these exceptional cases.
If numerical errors in a waveform are comparable to the small amplitude of the QNM,
    it will be hard to estimate $t_0$ and therefore the discrepancy will be large.
Estimation of $\omegaI$ is likely to be affected more sensitively than $\omegaR$ by the estimation error in $t_0$.
The average and the highest values of the relative errors over the waveforms are shown in Table~\ref{tab:relative_errors}.
It means that we can estimate the frequency and the damping rate of the QNM with sufficient accuracy
    if the QNM signal is exactly extracted from observed data.

\begin{table}[tbp]
    \centering
    \caption{
        Relative errors of $\omegaR$ and $\omegaI$ estimated for noise-free simulated waveforms.
    }
    \label{tab:relative_errors}
    \begin{tabularx}{.3\textwidth}{lRR}
        \hline \hline
            Relative errors [\%]    &   $\omegaR$   &   $\omegaI$   \\
        \hline
            average value           &   0.32        &   0.70        \\
            worst case              &   2.19        &   4.42        \\
        \hline \hline
    \end{tabularx}
\end{table}

\begin{figure*}[tbp]
    \vfill
        \hfill
        \subfigure[$\hat{t}_0$ histogram of 0305]{%
            \includegraphics{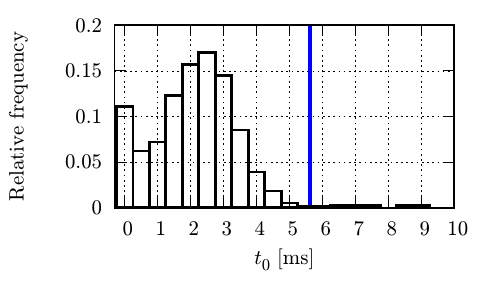}%
            \label{fig:0305_t0}%
        }
        \hfill
        \subfigure[$\hat{t}_0$ histogram of 0262]{%
            \includegraphics{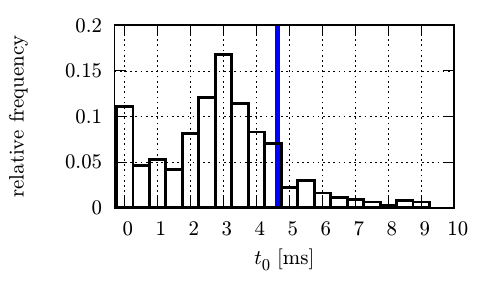}%
            \label{fig:0262_t0}%
        }
        \hfill
        \subfigure[$\hat{t}_0$ histogram of 0230]{%
            \includegraphics{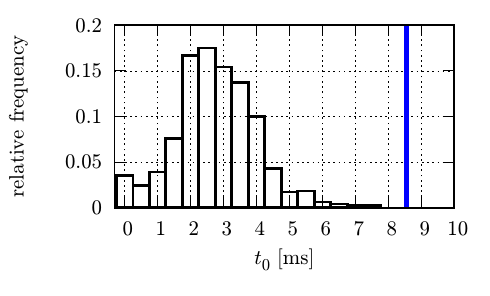}%
            \label{fig:0230_t0}%
        }
        \hfill
        \null
    \vfill
        \hfill
        \subfigure[$\hat{\omega}_\mathrm{R}$ histogram of 0305]{%
            \includegraphics{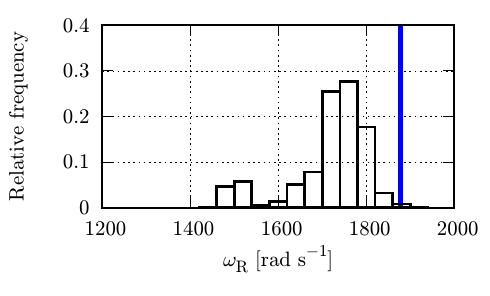}%
            \label{fig:0305_omegaR}%
        }
        \hfill
        \subfigure[$\hat{\omega}_\mathrm{R}$ histogram of 0262]{%
            \includegraphics{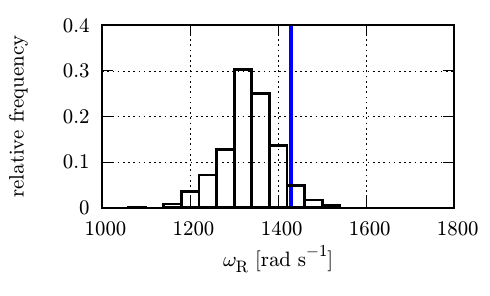}%
            \label{fig:0262_omegaR}%
        }
        \hfill
        \subfigure[$\hat{\omega}_\mathrm{R}$ histogram of 0230]{%
            \includegraphics{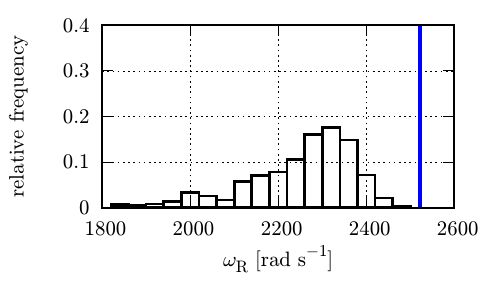}%
            \label{fig:0230_omegaR}%
        }
        \hfill
        \null
    \vfill
        \hfill
        \subfigure[$\hat{\omega}_\mathrm{I}$ histogram of 0305]{%
            \includegraphics{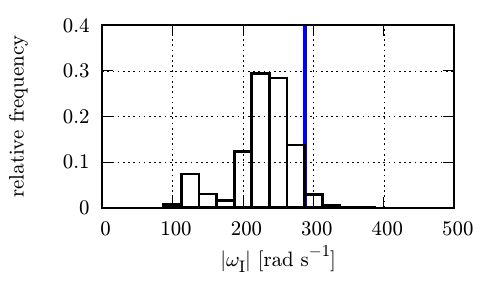}%
            \label{fig:0305_omegaI}%
        }
        \hfill
        \subfigure[$\hat{\omega}_\mathrm{I}$ histogram of 0262]{%
            \includegraphics{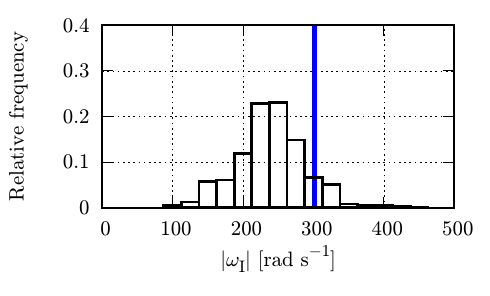}%
            \label{fig:0262_omegaI}%
        }
        \hfill
        \subfigure[$\hat{\omega}_\mathrm{I}$ histogram of 0230]{%
            \includegraphics{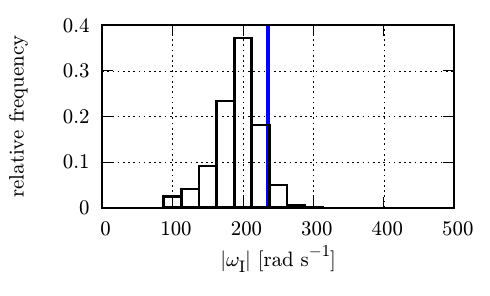}%
            \label{fig:0230_omegaI}%
        }
        \hfill
        \null
    \vfill
        \hfill
        \subfigure[$\hat{\omega}$ distribution of 0305]{%
            \includegraphics{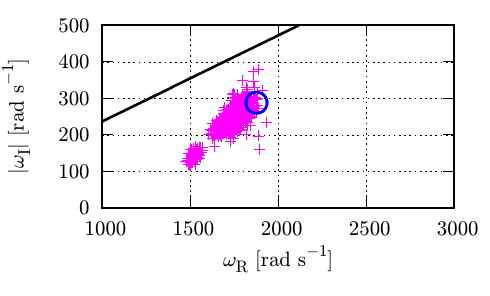}%
            \label{fig:0305_omega}%
        }
        \hfill
        \subfigure[$\hat{\omega}$ distribution of 0262]{%
            \includegraphics{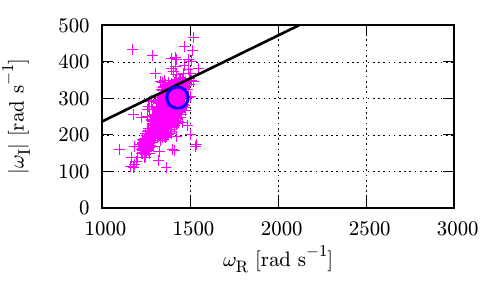}%
            \label{fig:0262_omega}%
        }
        \hfill
        \subfigure[$\hat{\omega}$ distribution of 0230]{%
            \includegraphics{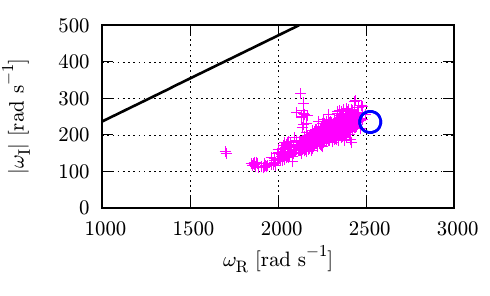}%
            \label{fig:0230_omega}%
        }
        \hfill
        \null
    \caption{
        Results of the analysis of QNM in simulated observation data for  waveform 0305, 0262 and 0230.
        Histograms of the estimated values of $t_0$, $\omegaR$, and $\omegaI$ are shown
            in the top row \subref{fig:0305_t0}--\subref{fig:0230_t0},
            the second row \subref{fig:0305_omegaR}--\subref{fig:0230_omegaR}
            and the third row \subref{fig:0305_omegaI}--\subref{fig:0230_omegaI} of the figure, respectively.
        The distributions of the estimated $\omegaR$ and $\omegaI$ are plotted
            in the bottom row \subref{fig:0305_omega}--\subref{fig:0230_omega}.
        The blue vertical line in each histogram shows
            the position along the horizontal axis for the value obtained through the analysis
                for the corresponding noise-free waveform,
                which is almost the same value with the theoretical value for $\omegaR$ and $\omegaI$.
        The black lines in panels \subref{fig:0305_omega}--\subref{fig:0230_omega}
            are the Schwarzschild limit and the area above this line is the prohibited area by GR.
        The centers of the blue circles in panels \subref{fig:0305_omega}--\subref{fig:0230_omega}
            indicate the values obtained for the corresponding noise-free waveforms.
    }
    \label{fig:results_of_simulated_strain}
\end{figure*}

To be more realistic,
    we make 1000 sets of ``simulated observation data'' for each waveform
        with simulated noise of Advanced LIGO's design sensitivity
    and apply our method to each set of them.
Some results of the analysis for three representative waveforms are shown in Fig.~\ref{fig:results_of_simulated_strain},
    and some parameters of these three waveforms are listed in Table.~\ref{tab:ParametersOfExamples},
        where an optimal SNR, $\rho$, for a waveform $h(t)$ and a detector sensitivity $S_\mathrm{n}(f)$ is given by
\begin{align}
    \rho = 2 \left( \int_{f_\mathrm{min}}^{f_\mathrm{max}} \frac{|\tilde{h}(f)|^2}{S_\mathrm{n}(f)} \rmd f \right)^{1/2}.
    \label{eq:optimalSNR}
\end{align}
We set $f_\mathrm{min}$ to $40$ Hz and $f_\mathrm{max}$ to $2048$ Hz,
    following Ref.~\cite{Allen_PRD_2012}.
Figures~\ref{fig:results_of_simulated_strain}\subref{fig:0305_t0}--\ref{fig:results_of_simulated_strain}\subref{fig:0230_omegaI}
    are the histograms of parameters $t_0$, $\omegaR$, $\omegaI$,
        obtained through the analysis,
    and \ref{fig:results_of_simulated_strain}\subref{fig:0305_omega}--\ref{fig:results_of_simulated_strain}\subref{fig:0230_omega}
        show the distributions of the estimated QNM frequencies.
The black solid line in each of these figures represents the Schwarzschild limit,
    and the area above this line is the prohibited area by GR\@.
The bin widths of the histograms are
    0.5 s for $t_0$, 100 rad/s for $\omegaR$, and 25 rad/s for $\omegaI$.
The gray-colored vertical line in each histogram shows the positions
    along the horizontal axis for the value
        obtained through the analysis for the corresponding noise-free waveform,
            which is almost the same value with the theoretical value for $\omegaR$ and $\omegaI$.
The centers of the gray circle in each distribution plot
    indicates the value obtained for the corresponding noise-free waveform.

\begin{table}[tbp]
    \centering
    \caption{
        Initial parameters of BBHs of waveforms 0305, 0262 and 0230.
        Optimal SNRs defined by Eq.~\eqref{eq:optimalSNR} are also shown.
        In this table,
            $\chi_{1z}$ and $\chi_{2z}$ respectively denote the initial spin components
            of the primary and the secondary BHs aligned with the orbital angular momentum of the BBH.
        For these three waveforms,
            either of BHs does not have the component of the initial spin perpendicular to the orbital angular momentum.
    }
    \label{tab:ParametersOfExamples}
    \begin{tabularx}{.4\textwidth}{lRRR}
        \hline \hline
            ID  &   0305    &   0262    &   0230    \\
        \hline
            mass ratio $q$  &
            $1.22$  &
            $3$     &
            $1$     
          \\
            initial spin $\chi_{1z}$  &
            $0.1$   &
            $-0.6$  &
            $0.8$   
          \\
            initial spin $\chi_{2z}$  &
            $-0.09$ &
            $0.0$   &
            $0.8$   
          \\
            SNR $\rho$  &
            $127.2$ &
            $83.07$ &
            $159.6$ 
          \\
        \hline \hline
    \end{tabularx}
\end{table}

The source parameters including the estimated starting times of the QNM for noise-free waveforms
    and the values obtained through our analysis for all 26 waveforms are listed in Table~\ref{tab:ParametersAndResults}.

\begin{turnpage}
\begin{table*}[tbp]
    \centering
    \caption{
        Parameters of source BBHs including $t_0$ estimated by applying our method to noise-free waveforms
            and the estimated values of $t_0$, $\omegaR$ and $\omegaI$ for the simulated observed data.
        From the first to the last column,
            the waveform ID,
            the initial mass ratio $m_1/m_2$,
            the $z$-components of the initial dimensionless spins of the primary and secondary BHs,
                $\chi_{1z}$ and $\chi_{2z}$, respectively,
            the ratio of the remnant mass $M_\mathrm{rem}$ to the initial total mass $M_\mathrm{t}$,
            the remnant dimensionless spin $\chi_\mathrm{rem}$,
            the real and imaginary parts of the theoretical value of the QNM frequency of the remnant BH,
                $\omegaR$ and $\omegaI$, respectively,
            the estimated value of the starting time of the QNM $t_0$
                for each of noise-free waveforms,
            the optimal SNR defined by Eq.~\eqref{eq:optimalSNR} for the simulated observed data,
                assuming $z = 0.09$, $M_\mathrm{t} = 60 M_\odot$ and Advanced LIGO's design sensitivity,
            and the estimated values of $\hat{t}_0$, $\hat{\omega}_\mathrm{R}$, $|\hat{\omega}_\mathrm{I}|$
            are listed.
        For each estimated value,
            the median value is shown with the region
                in which 68.2\% of the estimated values are included.
    }
    \label{tab:ParametersAndResults}
    \newcommand{\0}{\phantom{0}}
    \begin{tabularx}{.95\textheight}{rRRRRRRRRRRRRR}
        \hline \hline
            \\[-8pt]
            ID    & $m_1/m_2$ & $\chi_{1z}$ & $\chi_{2z}$
                        & $M_\mathrm{rem}/M_\mathrm{t}$ & $\chi_\mathrm{rem}$
                        & $\omegaR$/rad$\;$s$^{-1}$ & $|\omegaI|$/rad$\;$s$^{-1}$
                        & $t_0$/ms
                        & $\rho_\mathrm{IMR}$ & $\hat{t}_0$/ms
                        & $\hat{\omega}_\mathrm{R}$/rad$\;$s$^{-1}$
                        & $|\hat{\omega}_\mathrm{I}|$/rad$\;$s$^{-1}$
            \\
        \hline
            \\[-6pt]
            0209  &  1  &  $-0.90$  &  $-0.50$  &  0.965  &  0.46  &  1584  &  301  &  4.88  &  107.7
                     &  $2.2^{+1.2}_{-1.0}$  &  $1496^{+47.7\0}_{-56.3}$  &  $254^{+28.5}_{-27.7}$  \\[4pt]
            0210  &  1  &  $-0.90$  &  $+0.00$  &  0.961  &  0.54  &  1668  &  297  &  4.88  &  114.6
                     &  $2.2^{+1.7}_{-1.7}$  &  $1578^{+53.4\0}_{-97.7}$  &  $247^{+41.2}_{-42.4}$  \\[4pt]
            0211  &  1  &  $-0.90$  &  $+0.90$  &  0.951  &  0.68  &  1860  &  288  &  4.88  &  127.7
                     &  $2.2^{+1.7}_{-1.5}$  &  $1746^{+58.5\0}_{-120.8}$  &  $241^{+38.8}_{-44.6}$  \\[4pt]
            0212  &  1  &  $-0.80$  &  $-0.80$  &  0.967  &  0.43  &  1553  &  302  &  3.91  &  105.0
                     &  $2.4^{+1.7}_{-2.2}$  &  $1474^{+54.2\0}_{-105.0}$  &  $254^{+44.5}_{-65.4}$  \\[4pt]
            0218  &  1  &  $-0.50$  &  $+0.50$  &  0.952  &  0.69  &  1862  &  288  &  9.03  &  128.7
                     &  $2.2^{+1.7}_{-1.5}$  &  $1746^{+59.2\0}_{-103.6}$  &  $262^{+41.1}_{-51.5}$  \\[4pt]
            0221  &  1  &  $-0.40$  &  $+0.80$  &  0.945  &  0.74  &  1974  &  281  &  5.86  &  135.7
                     &  $2.4^{+1.2}_{-1.7}$  &  $1850^{+53.2\0}_{-117.6}$  &  $240^{+26.2}_{-38.3}$  \\[4pt]
            0223  &  1  &  $+0.30$  &  $+0.00$  &  0.947  &  0.73  &  1946  &  283  &  4.39  &  133.9
                     &  $2.2^{+1.7}_{-1.2}$  &  $1820^{+57.5\0}_{-133.2}$  &  $253^{+42.8}_{-47.8}$  \\[4pt]
            0230  &  1  &  $+0.80$  &  $+0.80$  &  0.911  &  0.91  &  2515  &  235  &  8.54  &  159.6
                     &  $2.9^{+1.2}_{-1.0}$  &  $2300^{+76.9\0}_{-143.4}$  &  $208^{+23.4}_{-32.1}$  \\[4pt]
            0231  &  1  &  $+0.90$  &  $+0.00$  &  0.935  &  0.82  &  2149  &  268  &  6.84  &  145.4
                     &  $2.4^{+1.5}_{-1.2}$  &  $1991^{+68.7\0}_{-127.0}$  &  $223^{+33.2}_{-33.1}$  \\[4pt]
            0232  &  1  &  $+0.90$  &  $+0.50$  &  0.920  &  0.88  &  2390  &  247  &  7.08  &  156.2
                     &  $2.9^{+1.2}_{-1.5}$  &  $2203^{+71.9\0}_{-167.5}$  &  $213^{+27.0}_{-43.9}$  \\[4pt]
            0259  &  2.5  &  $+0.00$  &  $+0.00$  &  0.967  &  0.58  &  1699  &  293  &  5.37  &  105.9
                     &  $2.0^{+1.7}_{-1.5}$  &  $1593^{+64.8\0}_{-91.8}$  &  $234^{+40.1}_{-43.7}$  \\[4pt]
            0262  &  3  &  $-0.60$  &  $+0.00$  &  0.978  &  0.27  &  1426  &  303  &  4.64  &  83.1
                     &  $3.2^{+1.5}_{-2.0}$  &  $1352^{+57.9\0}_{-50.0}$  &  $252^{+43.7}_{-43.8}$  \\[4pt]
            0263  &  3  &  $-0.60$  &  $+0.60$  &  0.977  &  0.29  &  1440  &  303  &  5.13  &  86.1
                     &  $2.7^{+1.7}_{-1.7}$  &  $1365^{+59.1\0}_{-46.1}$  &  $245^{+44.5}_{-39.7}$  \\[4pt]
            0264  &  3  &  $-0.60$  &  $-0.60$  &  0.979  &  0.25  &  1412  &  303  &  4.64  &  80.8
                     &  $3.2^{+1.5}_{-2.4}$  &  $1345^{+53.0\0}_{-77.5}$  &  $253^{+45.2}_{-63.2}$  \\[4pt]
            0265  &  3  &  $-0.60$  &  $-0.40$  &  0.979  &  0.26  &  1416  &  303  &  3.42  &  80.6
                     &  $2.7^{+1.7}_{-2.7}$  &  $1342^{+56.7\0}_{-86.7}$  &  $244^{+47.2}_{-63.8}$  \\[4pt]
            0274  &  3  &  $-0.23$  &  $+0.85$  &  0.973  &  0.47  &  1576  &  298  &  4.39  &  95.4
                     &  $2.7^{+2.0}_{-2.2}$  &  $1490^{+59.6\0}_{-83.7}$  &  $234^{+52.0}_{-51.5}$  \\[4pt]
            0276  &  3  &  $+0.30$  &  $+0.00$  &  0.972  &  0.53  &  1637  &  295  &  4.64  &  96.3
                     &  $2.4^{+1.5}_{-1.7}$  &  $1539^{+59.3\0}_{-60.0}$  &  $235^{+44.7}_{-37.5}$  \\[4pt]
            0283  &  3  &  $+0.30$  &  $+0.30$  &  0.965  &  0.68  &  1828  &  285  &  5.62  &  107.8
                     &  $2.7^{+1.2}_{-1.5}$  &  $1710^{+59.5\0}_{-85.3}$  &  $230^{+36.3}_{-40.4}$  \\[4pt]
            0287  &  3  &  $+0.60$  &  $-0.60$  &  0.960  &  0.78  &  2012  &  270  &  5.86  &  111.7
                     &  $2.2^{+1.5}_{-1.5}$  &  $1852^{+76.5\0}_{-136.5}$  &  $214^{+35.7}_{-43.9}$  \\[4pt]
            0289  &  3  &  $+0.60$  &  $+0.00$  &  0.957  &  0.80  &  2054  &  267  &  6.35  &  115.8
                     &  $2.2^{+1.5}_{-1.2}$  &  $1897^{+70.4\0}_{-139.5}$  &  $216^{+35.3}_{-42.2}$  \\[4pt]
            0292  &  3  &  $+0.73$  &  $-0.85$  &  0.956  &  0.83  &  2128  &  259  &  6.59  &  114.5
                     &  $2.2^{+1.5}_{-1.7}$  &  $1955^{+73.2\0}_{-171.8}$  &  $208^{+38.1}_{-42.7}$  \\[4pt]
            0293  &  3  &  $+0.85$  &  $+0.85$  &  0.936  &  0.91  &  2478  &  224  &  7.81  &  132.1
                     &  $2.4^{+2.4}_{-1.2}$  &  $2239^{+106.0}_{-216.5}$  &  $183^{+37.8}_{-49.4}$  \\[4pt]
            0297  &  6.5  &  $+0.00$  &  $+0.00$  &  0.987  &  0.35  &  1467  &  298  &  3.42  &  64.6
                     &  $2.7^{+1.7}_{-2.4}$  &  $1377^{+71.6\0}_{-82.3}$  &  $218^{+60.1}_{-57.4}$  \\[4pt]
            0303  &  10  &  $+0.00$  &  $+0.00$  &  0.992  &  0.26  &  1400  &  299  &  4.15  &  49.4
                     &  $2.2^{+2.0}_{-2.2}$  &  $1294^{+77.0\0}_{-77.6}$  &  $200^{+68.3}_{-55.4}$  \\[4pt]
            0305  &  1.22  &  $+0.10$  &  $-0.09$  &  0.952  &  0.69  &  1871  &  287  &  5.62  &  127.2
                     &  $2.4^{+1.2}_{-1.5}$  &  $1759^{+50.1\0}_{-92.0}$  &  $246^{+30.3}_{-34.6}$  \\[4pt]
            0317  &  3.33  &  $+0.30$  &  $-0.02$  &  0.964  &  0.75  &  1937  &  275  &  6.10  &  107.2
                     &  $2.0^{+1.5}_{-1.2}$  &  $1803^{+62.5\0}_{-95.0}$  &  $218^{+37.0}_{-34.4}$  \\[4pt]
    \hline \hline
    \end{tabularx}
\end{table*}
\end{turnpage}

\begin{figure}[htbp]
    \centering
    \includegraphics[clip]{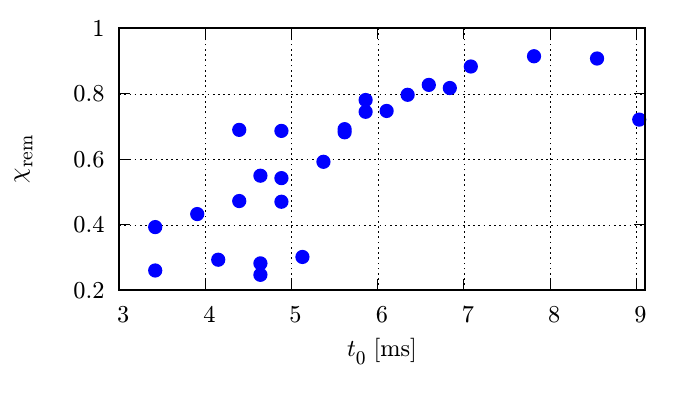}
    \caption{
        Spin of the remnant BH against the starting time of the QNM\@.
        The correlation coefficient $r$ is $0.764$.
    }
    \label{fig:correlation}
\end{figure}

We confirmed, for noise-free waveforms,
    that we can reasonably determine the QNM-dominant segment as shown by Fig.~\ref{fig:ideal_fits},
    and the QNM frequencies obtained in our analysis differ from theoretical values by only 1\% or less,
        although there are larger discrepancies in some cases,
        as shown by Fig.~\ref{fig:omega_comparison} and Table~\ref{tab:relative_errors}.
In Fig.~\ref{fig:correlation},
    we plot the spin of the remnant BH against the starting time of the QNM.
Apparently there is a correlation between the starting time and the remnant spin;
    correlation coefficient $r$ is 0.764.
It results from the fact
    that a large value of the spin causes 
        the merger phase to be long,
        and hence the starting time of the ringdown phase is delayed.
It is consistent with the analysis by Zhang \textit{et al.}~\cite{Zhang_PRD_2013}.

\begin{figure*}[htbp]
    \hfill
    \subfigure[Well-extracted case]{%
        \includegraphics{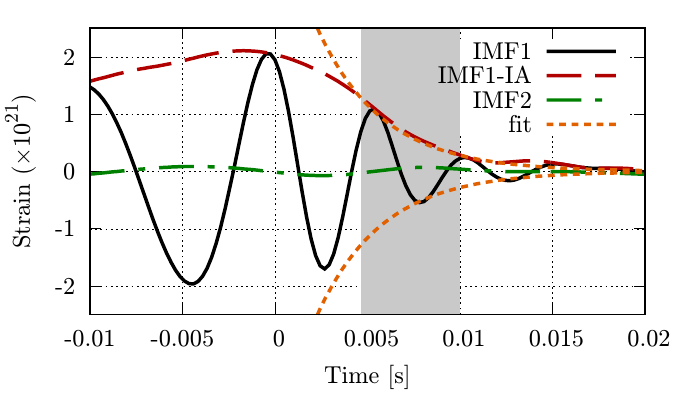}%
        \label{fig:0262_IMF_good}%
    }
    \hfill
    \subfigure[Frequent case]{%
        \includegraphics{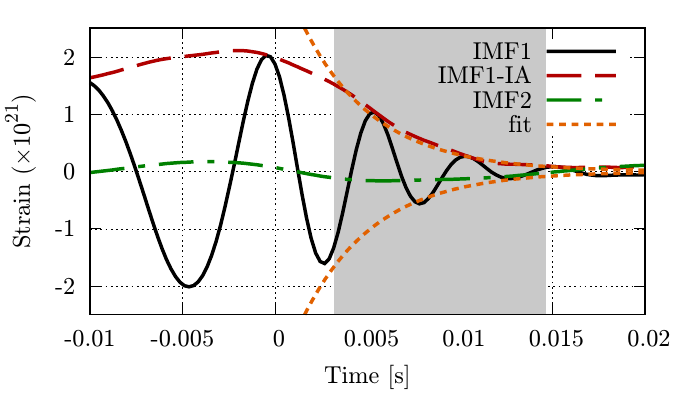}%
        \label{fig:0262_IMF_frequent}%
    }
    \hfill\null
    \caption{
        Two examples of the decomposition and fitting for waveform 0262
            with different time series of noise (series with different random number seeds).
        Panel~\subref{fig:0262_IMF_good} is the case
            that the estimate value of $t_0$ falls on the blue vertical line
            in Fig.~\ref{fig:results_of_simulated_strain}\subref{fig:0262_t0},
            while panel~\subref{fig:0262_IMF_frequent} is the case that $t_0$ falls in the most frequent bin.
        In each panel,
            IMF1 and IMF2 are drawn by the black solid line and the green dot-dashed line, respectively.
        The red dashed line, the orange dotted lines
            and the area filled with the color gray represent the same ones as Fig.~\ref{fig:ideal_fits}.
    }
    \label{fig:example_IMF}
\end{figure*}

However, the noise added to the waveform increases the estimation error of $t_0$ as shown
    by Figs.~\ref{fig:results_of_simulated_strain}\subref{fig:0305_t0}--\ref{fig:results_of_simulated_strain}\subref{fig:0230_t0}
    and Table~\ref{tab:ParametersAndResults}.
It tends to result in the smaller value of $t_0$,
    regardless of the SNR of a whole waveform and other parameters.
A primary source of this error is imperfectness of the mode decomposition.
In Fig.~\ref{fig:example_IMF},
    a comparison is drawn between two examples
        of the decomposition and fitting with the same waveform, 0262, but a different time series of noise.
The analysis with the data in Fig.~\ref{fig:example_IMF}\subref{fig:0262_IMF_good}
    yields the value of $t_0$ better than that in Fig.~\ref{fig:example_IMF}\subref{fig:0262_IMF_frequent},
    namely, 4.64 ms and 3.17 ms, respectively,
    while the value evaluated for the noise-free waveform is 4.64 ms.
We can observe that the amplitude of IMF2
    in Fig.~\ref{fig:example_IMF}\subref{fig:0262_IMF_frequent} is slightly higher than that
        in Fig.~\ref{fig:example_IMF}\subref{fig:0262_IMF_good}.
It means that the signal of the QNM is split into IMF1 and IMF2 in Fig.~\ref{fig:example_IMF}\subref{fig:0262_IMF_frequent},
    although that is almost perfectly extracted into IMF1 in Fig.~\ref{fig:example_IMF}\subref{fig:0262_IMF_good}.
This is caused by a drawback called ``mode-splitting'' of mode decomposition methods such as the EMD\@.

\section{Applying to GW150914}
\label{sec:GW150914}

\begin{figure*}[htbp]
    \hfill
    \subfigure[Raw data]{%
        \includegraphics{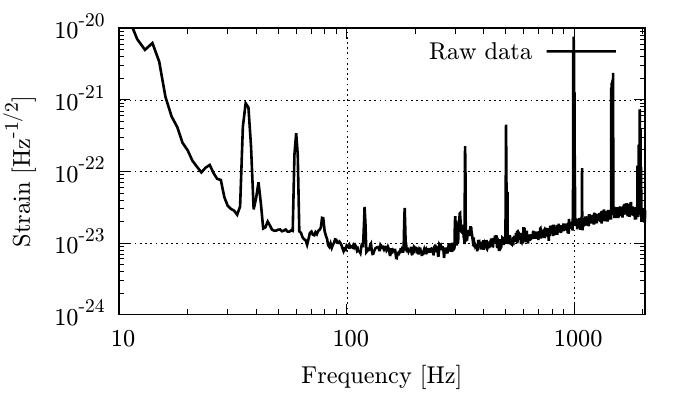}%
        \label{fig:ASD_GW150914_raw}%
    }
    \hfill
    \subfigure[Filtered data]{%
        \includegraphics{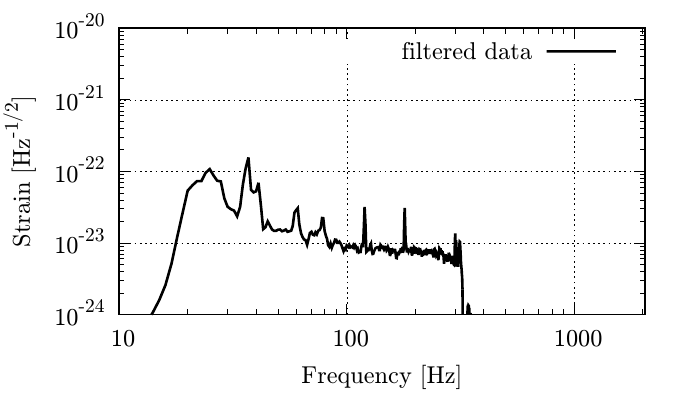}%
        \label{fig:ASD_GW150914_proc}%
    }
    \hfill\null
    \caption{
        Spectral strain sensitivity of observed data of GW150914.
        The left and right figures show the data before and after applying digital filters, respectively.
        The properties of the filter are listed in Table~\ref{tab:PropertyOfFilters}.
    }
    \label{fig:ASD_GW150914}
\end{figure*}

In this section, we apply our method to the strain data of GW150914
    from LIGO Hanford~\cite{LOSC_GW150914} to estimate its ability
    in analysis of real observed data.
The spectral strain sensitivity of the data is shown in Fig.~\ref{fig:ASD_GW150914}\subref{fig:ASD_GW150914_raw}.
Strong spectral lines are seen in the data.
To reduce large oscillation outside the most sensitive frequency band 
    and to attenuate the strong spectral lines inside it,
    we applied digital filters based on the Butterworth infinite impulse response filter of order 4.
The properties of the filters are listed in Table~\ref{tab:PropertyOfFilters},
    and the spectral strain sensitivity of the filtered data is shown
        in Fig.~\ref{fig:ASD_GW150914}\subref{fig:ASD_GW150914_proc}.

\begin{table}[tbp]
  \caption{
      Properties of filters we applied to the observed data of GW150914.
      These are digital filters based on the Butterworth infinite impulse response filter of order 4.
      In this table,
        $f_\mathrm{central}$, $\Delta f_\mathrm{pass}$ and $\Delta f_\mathrm{stop}$ denote
        the central frequency, pass bandwidth and stop bandwidth of a notch filter, respectively.
  }
  \label{tab:PropertyOfFilters}
  \centering
  \begin{tabularx}{.4\textwidth}{cRRR}
    \hline \hline
                & \multicolumn{3}{c}{Properties of notch filters [Hz]}
      \\
                & $f_\mathrm{central}$ & $\Delta f_\mathrm{pass}$ & $\Delta f_\mathrm{stop}$ 
      \\
    \hline
      Notch 1   &   36  &    1  &  0.1
      \\
      Notch 2   &   40  &    1  &  0.1
      \\
      Notch 3   &   60  &    1  &  0.1
      \\
      Notch 4   &  331  &   10  &    1
      \\
    \hline \hline
                & \multicolumn{3}{c}{Cutoff frequencies [Hz]}
    \\
                & &  Lower  &  Upper
    \\
    \hline
      Bandstop  & &  330.8  &  671.6 
      \\
      Bandpass  & &   20.0  &  320.0 
      \\
    \hline \hline
  \end{tabularx}
\end{table}

\begin{figure}[htbp]
    \centering
    \includegraphics{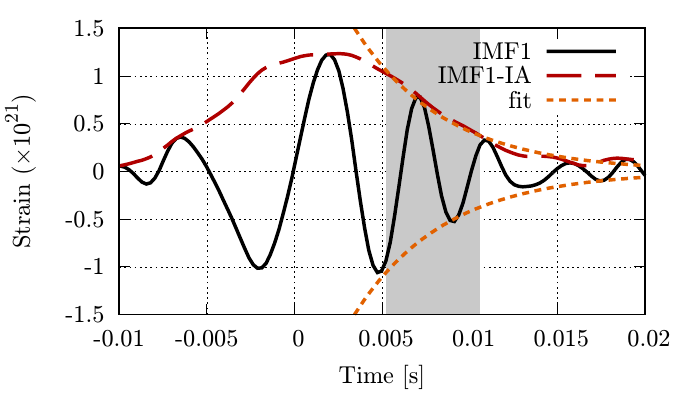}
    \caption{
        Fitting of the filtered data of GW150914.
        Lines and the gray-colored area represent the same ones as Fig.~\ref{fig:example_IMF}.
    }
    \label{fig:fitting_GW150914}
\end{figure}

Figure~\ref{fig:fitting_GW150914} illustrates the results of the analysis,
    which shows the time series data and the IA of the IMF1 obtained through the HHT analysis,
    the estimated QNM-dominant segment,
    and the exponentially decaying curve best-fitted to the IA\@.
It is similar to Fig.~\ref{fig:example_IMF}\subref{fig:0262_IMF_good},
    suggesting that the level of the accuracy in parameter estimates for real data is same as that for simulated data.
The evaluated values of parameters are listed in Table~\ref{tab:results_GW150914}.
The error ranges of $\omegaR$ and $\omegaI$ shown here are evaluated
    from only the standard deviation of the fitting of the IF and IA\@.
The corresponding frequency and damping time
    are $f = \omegaR/2\pi = 266 \pm 1.6$ Hz and $\tau = 1/|\omegaI| = 4.73 \pm 0.07$ ms, respectively.
Assuming that the redshift $z = 0.09$,
    and hence luminosity distance $d_\mathrm{L} = 410$ Mpc,
    following the parameter estimation in Ref.~\cite{LIGO_GW150914_estimation_PRL_2016},
    and that the QNM we extracted is of $(l,m,n) = (2,2,0)$,
    the dimensionless spin parameter $\chi_\mathrm{rem}$ and the mass $M_\mathrm{rem}$ of the remnant BH
    are evaluated from the QNM frequency $\omega$ by using the conversion table~\cite{Berti_PRD_2006,Berti_HP}.
They are shown in Table~\ref{tab:results_GW150914}, too,
    as well as measures of statistical errors corresponding to those of $\omegaR$ and $\omegaI$.
These values are apparently larger than those given by Ref.~\cite{LIGO_GW150914_testGR_2016}.
The reason is that, as shown  in Fig.~\ref{fig:results_of_simulated_strain},
    the systematic error due to noise in the observed data is large,
    and both $\omegaR$ and $\omegaI$ are tend to be underestimated.
So that the larger values of $\chi_\mathrm{rem}$ and $M_\mathrm{rem}$ are given.

\begin{table*}[htbp]
    \centering
    \caption{
        Estimated parameters of GW150914 by our method.
        The dimensionless spin parameter $\chi_\mathrm{rem}$ and the mass $M_\mathrm{rem}$ of the remnant BH
            are estimated from the QNM frequency $\omega = \omegaR + \mathrm{i}\omegaI$
            by using the conversion table~\cite{Berti_PRD_2006,Berti_HP},
                assuming the redshift $z = 0.09$ and the QNM mode of $l = m = 2$ and $n = 0$.
        The error ranges of $\omegaR$ and $\omegaI$
            are evaluated from only the standard deviation of the fitting of the IF and IA.
        The measures of errors of $\chi_\mathrm{rem}$ and $M_\mathrm{rem}$
            corresponding the errors of $\omegaR$ and $\omegaI$ are also shown.
    }
    \label{tab:results_GW150914}
    \begin{tabularx}{.75\textwidth}{RRRRR}
        \hline \hline
        \\[-8pt]
            $\hat{t}_0$/ms  &
            $\hat{\omega}_\mathrm{R}$/rad$\;$s$^{-1}$ & 
            $|\hat{\omega}_\mathrm{I}|$/rad$\;$s$^{-1}$ &
            \multicolumn{1}{c}{$\hat{\chi}_\mathrm{rem}$}   &
            $\hat{M}_\mathrm{rem}$/$M_\odot$
        \\[2pt]
        \hline
        \\[-7pt]
            $5.223$     &
            $1670 \pm 9.96$      &
            $211.2 \pm 3.13$     &
            $0.8151^{+0.092}_{-0.096}$    &
            $71.65^{+0.77}_{-0.75}$
            \\[3pt]
        \hline \hline
    \end{tabularx}
 \end{table*}

\section{Summary and Discussion}
\label{sec:summary}

We proposed a method of estimating the starting time of the QNM in the ringdown phase of GWs from BBH merger,
    and made its evaluation.
The determination of the starting time of the QNM is essential
    for the test of GR proposed by Nakano \textit{et al}.~\cite{Nakano_PRD_2015}.
We consider it as the commencing time of the segment
    in which the IA calculated with the HHT is well-fitted with an exponentially decaying
        form as it is expected for the QNM\@.
Also, we can estimate the QNM frequency of a remnant BH with this method.
The accuracy of the method was evaluated by applying it
    to noise-free gravitational waveforms free from detector noise provided by SXS group,
    simulated observation data,
    and then the real observed data of GW150914.
The simulated observation data are made by injecting waveforms
    into Gaussian noise based on Advanced LIGO's design sensitivity.

The results of analyzing noise-free waveforms show
    that the recovered QNM frequencies are consistent with theoretical values
    with accuracy of 0.32 \% and 0.70 \%, on average,
    for the real and imaginary parts, respectively.
This means that
    we can recover parameters of the remnant BH correctly
    ONLY if the QNM signal is perfectly extracted.
In future,
    third generation GW detectors, such as the Einstein Telescope~\cite{ET_CQG_2010}
    and B-DECIGO~\cite{B-DECIGO_PTEP_2016}
    will observe BBH signals with enormously high SNR.
With these detectors,
    our method can work properly and contribute to some tests of GR using QNMs.
We find also
    that starting times of the QNM have a correlation with spins of remnant BHs.
It is attributed to the fact that the merger phase remain longer,
    and hence the starting time of the QNM becomes later as the spin of the remnant BH is larger.

At present,
    the efficiency of our method for observed data is not so high.
It is mainly caused by the fact that the EMD is still in developing stage.
Noise in observed data will often give rise to the mode-splitting in the EMD used in the HHT\@.
In this case,
    the peak value of the IMF is reduced,
    and therefore the signal will be observed to decay more slowly.
It also causes the amplitude of the IMF to fit from earlier to an exponentially decaying curve.
It results in the smaller value of the imaginary part of the QNM frequency
    and the smaller value of the starting time of the QNM\@.
The deformation of the IMF due to noise usually leads to the smaller value of the IF
    through the Hilbert spectral analysis,
    and thus the real part of the QNM will be underestimated.

Even though several improvements of the EMD have been proposed,
    the mode-splitting problem has not yet been resolved.
If the mode-splitting is controlled more effectively
    and thus the signal of the QNM is extracted into a single IMF,
    the parameters will be estimated more accurately through our method.
One promising solution is to take sparsity
    in the frequency domain into account in the mode decomposition.
This idea is based on the fact that the instantaneous frequency of the QNM is constant.
It should be noted,
    however, that this solution can be applied to only some specific cases,
        in which the frequency of the mode of interest is known to be constant,
    and that it will not always resolve the mode-splitting problem.

We are now focusing on the only dominant, slowest-damped mode $(l,m,n) = (2,2,0)$,
    by filtering higher modes such as $(l,m) = (3,3)$ modes.
According to London, \textit{et al.}~\cite{London_PRD_2014},
    although $(l,m,n) = (2,2,1)$ and $(l,m,n) = (2,1,0)$ modes are in the same band of the target mode,
    their amplitudes are 1 order of magnitude lower than the target mode.
We have not considered the contributions of these modes in this paper.
For further work,
    we are planning to analyze mixtures of various modes.
Improvement of mode decomposition will enhance the ability
    to decompose different modes into different IMFs properly,
    and therefore we will be able to investigate higher QNMs,
    as well as the fundamental mode, accurately.

\section*{Acknowledgement}

This research made use of data obtained from the LIGO Open Science Center~\cite{LOSC_HP},
    a service of LIGO Laboratory and the LIGO Scientific Collaboration,
    and made use of data generated by the Simulating eXtreme Spacetimes.
LIGO is funded by the U.S. National Science Foundation.
This work was in part supported by MEXT Grant-in-Aid for Scientific Research 
    on Innovative Areas ``New Developments in Astrophysics Through Multi-Messenger Observations of Gravitational Wave Sources''
    (Grants No.~24103005 and No.~24103006)
    and ``Gravitational Wave Physics and Astronomy: Genesis'' (Grants No.~17H06357 and No.~17H06358).
This work was also supported in part
    by the joint research program of the Institute for Cosmic Ray Research, University of Tokyo (K.S., K.O., M.K. and H.T.),
    JSPS Grant-in-Aid for Scientific Research (C)
    (Grant No.~15K05071; K.O., Grant No.~16K05347; H.N., Grant No.~17K05437; H.T.)
    and JSPS Grant-in-Aid for Young Scientists (B) (Grant No.~26800129; H.T.).
We would like to thank Nami Uchikata for carefully reading our manuscript.

\bibliographystyle{apsrev4-1}
\bibliography{main}

\end{document}